# The Thermally Reversing Window in Ternary $Ge_xP_xS_{1-2x}$ glasses

U. Vempati and P. Boolchand
Department of Electrical, Computer Engineering and Computer Science
University of Cincinnati, Cincinnati, Ohio 45221-0030

## Abstract

$Ge_xP_xS_{1-2x}$ glasses in the compositional range $0.05 \leq x \leq 0.19$ have been synthesized and examined in temperature modulated differential scanning calorimetry (MDSC) and Raman scattering experiments. Trends in the non-reversing enthalpy $\Delta H_{nr}(x)$ near $T_g$ show the term to almost vanish in the $0.090(5) < x < 0.135(5)$ range, and to increase by an order of magnitude at $x < 0.09$, and at $x > 0.135$. In analogy to previous results on chalcogenide glasses, we identify compositions at $x < 0.09$ to be elastically *floppy*, those in the $0.090 < x < 0.135$ range to be in the *intermediate phase*, and those at $x > 0.135$ to be *stressed rigid*. MDSC results also show the $\Delta H_{nr}$ term *ages* in the stressed-rigid and floppy phases but *not* in the intermediate phase. The intermediate phase is viewed to be a self-organized phase of a disordered network. It consists of at least four isostatically rigid local structures; corner-sharing $GeS_4$, edge-sharing $GeS_2$, pyramidal $P(S_{1/2})_3$ and quasi-tetrahedral $S=P(S_{1/2})_3$ units for which evidence comes from Raman scattering. The latter method also shows existence of $P_4S_7$ and $P_4S_{10}$ molecules in the glasses segregated from the backbone. These aspects of structure contribute to an intermediate phase that is significantly narrower in width than in corresponding selenide glasses.

# 1. Introduction

Under certain conditions disordered networks can self-organize. The recognition emerged [1] from observation on binary $Si_xSe_{1-x}$ glasses of *two* elastic phase transitions as a function of network connectedness. In Raman scattering, the optical elasticity displays two thresholds, a second-order threshold near $x = x_c(1) = 0.20$ and a first order one near $x = x_c(2) = 0.27$. In MDSC measurements, the non-reversing enthalpy ($\Delta H_{nr}$) near $T_g$ displays an almost square-well like behavior with the walls located near $x_c(1)$ and $x_c(2)$, and with the heat-flow term nearly vanishing in the well, $x_c(1) < x < x_c(2)$. Such thermally reversing windows that coincide with Raman optical elasticity thresholds have now been observed [2-5] in other chalcogenide glass systems underscoring the generality of the observation.

A solitary elastic phase transition in random networks from an elastically floppy to stressed-rigid state was predicted by J.C. Phillips [6] and M.F. Thorpe [7] in the early 1980s. An important step in understanding the origin of two elastic phase transitions came from the numerical simulations of M.F. Thorpe et al. [8] who suggested the first transition represents the onset of rigidity ($x_c(1)$) and the second transition the onset of stress ($x_c(2)$) in the backbone of these glasses. The intervening region, the intermediate phase, represents then a rigid but stress-free domain of a disordered network identified with the *self-organized* phase of a glass. M. Micoulaut and J.C. Phillips [9] have more recently introduced an analytical method called <u>s</u>ize <u>i</u>ncreasing <u>c</u>luster <u>a</u>gglomeration (SICA) to estimate the compositional range of the intermediate phase in Si-Se glasses. The method starts from the appropriate local structures, and these are then agglomerated in steps to build medium range structures. Constraint counting algorithms [9] then permit establishing the chemical compositions where rigidity onsets $x_c(1)$, and then stress onsets $x_c(2)$ in these clusters as they are cross-linked progressively. The intermediate phase estimated by SICA is in good agreement with the experimental results on Si-Se and Ge-Se glasses. SICA calculations become progressively more intensive as the number of local structures and agglomeration steps are increased.

In this work we examine the nature of elastic phases in ternary $Ge_xP_xS_{1-2x}$ glasses. Glass structure is studied using Raman scattering, and the thermal properties of the glasses including $T_g$ examined in MDSC measurements. These results permit a detailed comparison of the intermediate phase in the present glasses with that in corresponding selenides [5]. The comparison provides new insights into aspects of molecular structure that control the width of the self-organized phase when Se is replaced by S. Self-organization effects in disordered systems represent a recent and exciting issue in glass science that has profound consequences in a number of fields including protein folding [10], thin-film gate dielectrics [11] for CMOS devices, and the origin of high temperature superconductivity [12]. The subject is close to the continuing interests [13] of M.F. Thorpe, and it is, indeed, a pleasure to dedicate this contribution to celebrate his 60[th] anniversary.

## 2. Experimental

### 2.1. Sample Synthesis

99.999% elemental Ge, S and P from Cerac Inc., were used as starting materials to synthesize ternary $Ge_xP_xS_{1-2x}$ glasses in the compositional range $0.05 < x < 0.19$. One gram batches were sealed in evacuated (5 x 10$^{-7}$ Torr) quartz ampoules and slowly (2 days) heated to 950°C. Melts were homogenized for 3 days in a rocking furnace. They were then equilibrated 30°C to 50°C above the liquidus for 6 hours before a water quench to obtain the glasses. At low-x (< 3%) and high-x (>24%), we were unable to obtain uniform glasses. Once synthesized the samples were handled in a dry nitrogen ambient and stored in evacuated Pyrex tubings.

*2.2. MDSC Results*

A model 2920 MDSC from TA instruments Inc, operated at 3°C min$^{-1}$ scan rate and 1°C / 100s modulation rate was used to examine the glass transitions [14]. Samples aged at room temperature for 2 weeks or longer, were scanned up in temperature past $T_g$ and then cooled back to room temperature. $T_g$ was deduced from the inflexion point of the reversing heat flow signal. At x = 0.05, one observes, in general, two endotherms in the glasses (Fig. 1a) one near 55°C and the other near 150°C. Here, the low-T endotherm represents the glass transition temperature while the high-T endotherm the polymerization [15-16] transition $T_\lambda$ of sulfur. With increasing x, the low-T endotherm moves up in T while the high-T endotherm remains nearly fixed near T = 150°C.

A characteristic feature of the $T_\lambda$ transition is the unusual non-reversing heat flow associated with it; an exotherm followed by an endotherm. It appears that $S_8$ crowns relax towards the backbone (exotherm) prior to opening into $S_n$-chain fragments (endotherm). In orthorhombic S, where there is no network backbone, the $T_\lambda$ transition [16] consists of an endotherm alone. In the present glasses the normalized integrated area under the $T_\lambda$-endotherm decreases with x, and it reflects the concentration of the $S_8$ crowns decreasing and eventually vanishing near x = 0.09 (Fig. 2).

The non-reversing heat flow extracted from the difference of the *total*- and *reversing*- heat flow signals scanning up in temperature suffers from an overestimate due to an upshift of the latter signal because of the finite scan rate employed (3ºC/min). To correct [17] for this shift, one scans down in temperature across $T_g$ after scanning up. One then obtains the corrected $\Delta H_{nr}$ term by taking the difference between the non-reversing heat flow scanning up in T ($\Delta H_{nr}^{up}$) from the non-reversing heat flow coming down in T ($\Delta H_{nr}^{down}$) as shown in Fig.1b. In the latter scan, the reversing heat flow signal downshifts in T as much as it upshifted in the scan up in T. Henceforth, the corrected non-reversing heat flow term will simply be referred as $\Delta H_{nr}$. For the glass composition at x = 0.11, one can see from Fig.1b that the $\Delta H_{nr}$ term of 0.02 cal/gm is minuscule.

Fig. 3a gives a summary of the compositional variation in $T_g(x)$ of the glasses. We find $T_g(x)$ to increase monotonically with x, at first sharply in the 0.05 < x < 0.10 range and thereafter more slowly once x > 0.10. Fig. 3a also shows glass compositions across which the $T_\lambda$ transition is observed. Compositional trends in $\Delta H_{nr}(x)$ are summarized in Fig. 3b in which, we display results on glass samples aged for 2 weeks after melt-quenching, and separately for 8 weeks. The results of Fig. 3a bring out the exceptional behavior of glass compositions in the 0.09 < x < 0.135 range; these glasses possess a vanishing $\Delta H_{nr}$ term and that term does not age. On the other hand, glass compositions at x < 0.09 and at x > 0.135 not only possess an order of magnitude larger $\Delta H_{nr}$ term but that term also appears to age noticeably.

2.3. *Raman Scattering*

Raman scattering was excited using 1mW of 514.5 nm radiation loosely focused to a 50 μm laser spot size on glass sample encapsulated in quartz tubings, which were used to synthesize the samples. The back-scattered radiation was analyzed using a model T-64000 triple-monochromator system [4,16] from Jobin Yvon Inc. using a charge coupled device detector. Fig 4a gives an overview of the observed Raman line shapes at several glass compositions. At low x (5% - 9%), the line shapes are dominated by modes [15,16] of elemental S at 150 cm$^{-1}$, 215 cm$^{-1}$ and 485 cm$^{-1}$ that steadily decrease in scattering strength as x increases. Concomitantly, one observes an increase in scattering strength of several modes; corner-sharing (CS) GeS$_4$ tetrahedra at 345 cm$^{-1}$, edge-sharing (ES) Ge(S$_{1/2}$)$_4$ tetrahedra at 375 cm$^{-1}$, pyramidal (PYR) P(S$_{1/2}$)$_3$ units at 416 cm$^{-1}$ and quasi-tetrahedral (QT) S=P (S$_{1/2}$)$_3$ tetrahedra near 700 cm$^{-1}$. At higher x, sharp modes of P$_4$S$_{10}$ and P$_4$S$_7$ monomers appear in the line shape. In the spectra, we can also observe sharp modes of the P$_4$S$_4$ and P$_4$S$_3$ monomers (containing P-P bonds) at x > 0.15 highlighted in the inset of Fig. 4a. An enlarged view of the two insets of Fig. 4a appears as Fig. 4b and 4c in which the mode assignments of the different species are indicated.

**Table 1** gives a summary of the vibrational mode assignments in the present glasses. These assignments were made feasible by earlier vibrational studies on binary Ge-S glasses [18,19], P-S glasses [20-22], crystalline molecular solids [23-25] in the P-S binary, amorphous P [26-28] as well as first-principles cluster calculations [29,30]. Figure 5 provides a schematic drawing of these local structures, some of which form part of the backbone while others segregate as molecules in the glasses. We shall return to discuss these results in section 3.

The observed line shapes were least-squares fit to the requisite number of Gaussians with unrestricted line-width, centroid and intensity. An example is provided in fig 4(d) that provides line shape analysis of the low frequency segment of a glass sample at x = 0.15. One can clearly identify vibrational modes of characteristic monomers (Table 1). In our analysis of these line shapes we have used a software package, Peak-Fit, marketed by SPSS Inc. to deconvolute the line shapes. The normalized scattering strength of the 700 cm$^{-1}$ mode (QT units), 345 cm$^{-1}$ mode (CS units), 270 cm$^{-1}$ mode ($P_4S_{10}$ units), 180 cm$^{-1}$ mode ($P_4S_3$ units), 230 cm$^{-1}$ mode ($P_4S_7$ units) are summarized in Fig.6. On this plot the scattering strength of the indicated modes were normalized to the total area. No attempts to obtain Raman cross sections of the mode were made. Until these matrix element effects are folded in, Raman scattering strengths can serve only as a qualitative guide of local structure concentrations.

## 3. Discussion

*3.1. Evolution of glass structure with chemical composition*

*3.1.1 Low cation concentration x (< 0.10)*

Sulfur melts upon cooling form molecular crystals [16] composed of $S_8$ crowns. The behavior is unlike that of Selenium melts that give rise to a glass composed of polymeric $Se_n$ - chain fragments upon cooling. There have been efforts to rapidly quench sulfur melts to form bulk glasses, and a $T_g$ of about 255K or - 18ºC has been reported [31]. The low $T_g$ is suggestive of a zero dimensional glass composed largely of $S_8$ monomers coupled by weaker van der Waals forces. In general, it is difficult to form homogeneous S-rich bulk glasses with trace amounts of group IV and V additives largely because of the tendency of $S_8$ crowns to segregate in such samples. At room temperature, these monomers are chemically more stable than $S_n$-chains.

On chemical grounds one may expect P and Ge additives in a sulfur melt to result in formation of heteropolar bonds because the respective single P-S (54.8 kcal/mole) and Ge-S (55.5 kcal/mole) bond strengths [32] (Table 2) exceed the homopolar bond strengths of S-S (50.9 kcal/mole), Ge-Ge (37.6 kcal/mole) and P-P (51.3 kcal/mole). Specific P-centered and Ge-centered local structures consistent with the chemical valence of P- and Ge- can be expected to form in the present ternary glasses as they do in corresponding binary (Ge-S, P-S) glasses [18-22]. The prevailing P-centered and Ge-centered local structures populated in the present glasses include PYR [P($S_{1/2}$)$_3$], QT [S=P($S_{1/2}$)$_3$], CS [GeS$_4$] and ES [GeS$_2$] for which Raman vibrational evidence is available from our results (Fig. 4 and Table 1).

The molecular structure of S-rich glasses is thus expected to be intrinsically heterogeneous and to consist of $S_8$ crowns and a network backbone composed of specific P- and Ge- based local structures serving to crosslink $S_n$- chains. With increasing cation concentration x, one expects the network backbone to grow at the expense of the $S_8$ monomer fraction. The former structural grouping gives rise to a glass transition temperature, $T_g$, while the latter monomers to the $T_\lambda$ transition as observed in our MDSC scans (Fig. 1a). Our experiments show that as x increases to 0.09, the $S_8$ monomer fraction becomes minuscule (Fig.2).

*3.1.2. High cation concentration x (> 0.10)*

At higher x (10 % < x < 15%), more tetrahedral units (GeS$_4$), pyramidal units (P($S_{1/2}$)$_3$) and quasi-tetrahedral units (S=P($S_{1/2}$)$_3$) appear in the backbone as revealed by an increase in scattering strengths of their vibrational modes (Fig. 6). In addition to these network forming local structures, our Raman scattering results also reveal monomers of P$_4$S$_{10}$ and P$_4$S$_7$ to systematically grow in concentration with increasing x (Fig. 4 and Fig.6). The P$_4$S$_{10}$ monomers first appear in the glasses at x = 0.05, and their concentration steadily increases to peak near x = 0.13. These units persist (Fig.6) across the intermediate phase (0.09 < x < 0.135). The P$_4$S$_{10}$ monomers convert to P$_4$S$_7$ ones, their P-richer counterparts, in the $0.15 < x < 0.19$ range as S is depleted from the glasses. The view is corroborated from the low frequency (Fig.4b) and high frequency modes (Fig. 4c) of the $P_4S_7$ species that increase in strength with x.

In the present ternary glasses, one expects a chemical threshold [2] to occur when x increases to $x_t = 2/11$. If we require P to be 3-fold and Ge to be 4-fold coordinated to S, then on stoichiometric grounds one must require

$$P_xGe_xS_{1-2x} = [P(S_{1/2})_3]_x [Ge(S_{1/2})_4]_x [S_{1-2x-7x/2}] \qquad (1)$$

A chemical threshold will appear when there is not enough S per P and Ge atom to permit pyramidal $P(S_{1/2})_3$ units and tetrahedral $Ge(S_{1/2})_4$ units to be formed in the network. The condition will occur when the free S content in equation (1) vanishes, i.e.,

$$1 - 2x_t - 7x_t/2 = 0, \text{ or } x_t = 2/11 \qquad (2)$$

Homopolar (Ge-Ge, P-P) bonds can be expected to emerge at $x > x_t$. However, since P-P bonds (51.3 kcal/mole) are more strongly bound than Ge-Ge bonds (37.6 kcal/mole, Table 2), one expects P-P bonds to appear in the network at $x < x_t$, and thereby delaying the appearance of Ge-Ge bonds to a value of $x > x_t$. The appearance of the latter bonds in the glasses can be established using $^{119}$Sn Mossbauer spectroscopy as recently demonstrated [2] in ternary $Ge_xAs_xSe_{1-2x}$ glasses. Such experiments are currently ongoing and will be reported upon later.

In the Raman spectra, the appearance of vibrational modes of $P_4S_4$ and $P_4S_3$ monomers (contain P-P bonds, Fig. 5) at x > 0.15 is thus in harmony with these chemical considerations. $P_4S_{10}$ and $P_4S_7$ monomers are terminated by singly coordinated P=S species. These monomers occur with the S anion (but not with Se anion), and are formed in the glasses starting from x = 0.05 to steadily increase in concentration with x to peak in the stressed-rigid glass compositions. And as one would expect on simple bond statistics [33], the S-rich monomer ($P_4S_{10}$) peaks (Fig. 6) at a lower x (= 0.13) than the S-deficient ($P_4S_7$) one (x = 0.18).

As more P- and Ge- is alloyed in the glasses, one expects network fragments of amorphous P and amorphous Ge to emerge as was recently observed in an examination of ternary $Ge_xP_xSe_{1-2x}$ glasses at x > 0.20 by S. Chakravarty et al. [34]. Vibrational signatures of tetrahedrally coordinated $P_4$ clusters [27] (460 $cm^{-1}$, 590 $cm^{-1}$) and amorphous P [26] network (390 $cm^{-1}$, 490 $cm^{-1}$) were observed in the selenides. In the present ternary, we do not have enough glass compositions at x > 0.17 to meaningfully elucidate these phases, however, we do observe a mode at 470 $cm^{-1}$ in a sample at x = 0.19 which we assign to the presence of tetrahedral $P_4$ clusters in the present sulfides.

### 3.1.3. Glass transition temperature of a sulfur-chain glass.

The magnitude of $T_g$ contains vital structure information. The recognition emerged with the inception of stochastic agglomeration theory by R. Kerner and M. Micoulaut [35,36]. In binary $(T)_xSe_{1-x}$ glasses, where T = Si or Ge, the variation of $T_g$ with x is predicted [36] to be linear with a slope, $dT_g/dx$ of $T_o$/ln 4/2, that is independent of x in the stochastic limit (low x ~ 5%). Here $T_o$ represents the $T_g$ of Se glass, and the numbers of 4 and 2 represent the coordinations of Ge or Si and Se respectively. This prediction is in remarkable agreement [37] with experiments, and serves to underscore the notion that the magnitude of $T_g$ serves as a good representation of network connectedness or mean coordination number $\bar{r}$. It is becoming increasingly clear that $\bar{r}$ is an important topological factor necessary to understanding the physical behavior [6] of disordered systems. The importance of chemical bond strengths in this respect is to serve as a scaling factor [38] in comparing networks having the same connectedness or mean $\bar{r}$.

In the present ternary sulfides, the increasing value of the glass transition temperature $T_g(x)$ with x constitutes signature of an increase in network connectedness as P and Ge atoms cross-link $S_n$ chains. The low value of $T_g$ at x = 0 is most likely due to the absence of a network in the case of elemental S as mentioned earlier. The rapid increase of $T_g(x)$ at x > 0.05, is most likely due to the formation of a covalently bonded network of $S_n$ chains mildly cross-linked by the P- and Ge additives as the $S_8$ crown concentration vanishes (Fig.2) near x = 0.10.

It is instructive to compare the $T_g(x)$ variation in the present ternary sulfides with corresponding selenides [34] as illustrated in Fig. 7. In the 0.09 < x < 0.16 range, one finds the variations in $T_g(x)$ for *both* ternaries (Fig 7) to be strikingly parallel with a scaling factor of 1.13. For example, at x = 0.14, we obtain a $T_g$ in the ternary sulfide glass of 241 ºC or 514K, and in a corresponding ternary selenide glass of 180 ºC or 453K, yielding a scaling factor of 514K/453K = 1.13(1). The scaling reflects *similarities* of local structures of the backbones in the two chalcogenides in the 0.10 < x < 0.16 range. The scaling factor can be traced to the higher bond strengths of Ge-S and P-S bonds in relation to Ge-Se and P-Se bonds. The Pauling single bond strength ($E_b$) listed in Table 2, reveal that the ratio

$$[E_b(\text{Ge-S}) + E_b(\text{P-S})] / [E_b(\text{Ge-Se}) + E_b(\text{P-Se})] = 1.10 \qquad (3)$$

We thus find that the chemical bond strength scaling factor from equation (3) is in remarkable agreement with the $T_g$ scaling factor of the glasses. These correlations serve to reinforce the topological or structural interpretation of $T_g$ [37,38] in network forming materials rather than the strictly chemical one advanced earlier [39].

The extrapolated value of $T_g$ at $x = 0$ as shown by the dashed line in Fig. 7 in the present sulfide glasses gives a value of 115º C. This temperature then represents the predicted value of $T_g$ of a $S_n$-chain glass based on the present work. Unfortunately, it is too close to the melting temperature of elemental sulfur and may well represent the basic limitation to form a $S_n$-chain glass.

3.2. *Elastic Phases in $Ge_xP_xS_{1-2x}$ glasses*

The central result of the present study is the discovery of a *thermally reversing window* in the $0.09 < x < 0.135$ (Fig. 2) composition range. The result follows a pattern noted earlier [16, 40] in other chalcogenide glasses. In analogy to previous studies [8, 9,16,40], we identify glasses at $x < 0.09$ to be elastically *floppy*, those in the $0.09 < x < 0.135$ range to be in *the intermediate phase*, and those at $x > 0.135$ to be *stressed rigid*. Some of the physical characteristics of these elastic phases suggested from the present observations are discussed in the following discussion.

3.2.1. *Elastically Floppy Phase ($x < 0.09$)*

The molecular structure of glasses at low x consists of $S_8$ crowns that segregate from a backbone of $S_n$ chain fragments crosslinked by P-centered and Ge-centered local structures. The concentration of $S_8$ crowns is high at x = 0.05, but it steadily declines as x increases to 0.09. The non-reversing enthalpy of glasses in the floppy phase, $\Delta H_{nr}$, displays a nearly symmetric peak that is about 25º C wide (Fig.1a). Upon aging the $\Delta H_{nr}$ term increases by an order of magnitude (Fig.1a), a feature that is tied to the mechanically under-constrained nature of $S_n$ chains (Table 3). In a chain, a S atom has one bond-stretching and one bond-bending constraints. Since the total count of constraints per sulfur atom of 2 is less than 3, the dimensionality of the network, $1/3^{rd}$ of the vibrational modes are elastically floppy. Upon aging the chains flop around increasing the entropy of the network reflected in an increase of $\Delta H_{nr}$. Fig.1a also reveals that a sample at x = 0.05, ages much less than one at x = 0.08, we suppose because the fraction of S atoms present in the backbone in the former glass is minuscule in relation to that in the later glass. These results reinforce that the non-reversing enthalpy near $T_g$, $\Delta H_{nr}$, is a backbone related property.

A comparison of the $\Delta H_{nr}$ term in Ge-P-Se and Ge-P-S glasses (Fig. 7) shows that at x < 0.08 the term ages much more in the selenides than in the sulfide glasses. Se-rich glasses are, in general, more polymerized than their S-rich counterparts as alluded to earlier. The presence of pronounced aging effects in the selenide glasses but its near absence in the sulfide glasses at low x, underscores that the $\Delta H_{nr}$ term is intrinsically a backbone related property. In particular, monomeric units like $S_8$ or $P_4S_{10}$ that are decoupled from the network backbone apparently do not contribute to the non-reversing heat flow term.

*3.2.2. Intermediate Phase (0.090 < x < 0.135)*

The conditions leading a glass network to self-organize are very selective. A network must consist of local- and medium- range structures that are rigid but with no redundant bonds, i.e., it must be *isostatically rigid*. The presence of redundant bonds gives rise to stress and an increase in free energy. A network can lower its free energy by avoiding redundant bonds and forming a globally isostatically rigid backbone.

A covalent bond between two atoms provides a *bond-stretching* constraint that is shared equally between the two atoms. Thus, an atom having $\bar{r}$ nearest-neighbors will have $\bar{r}/2$ bond-stretching constraints. In addition, there are $2\bar{r}-3$ independent bond angles (in 3d) that must be specified to fix the local geometry of an atom having $\bar{r}$ ($\geq 2$) nearest neighbors, leading then to an equivalent number of *bond-bending* constraints. The total count of bonding constraints [6,7] then equals ($5\bar{r}/2 - 3$), i.e., 7 for Ge ($\bar{r}=4$), 2 for S ($\bar{r}=2$) and 4.5 for P ($\bar{r}=3$). For the special case of an atom having $\bar{r}=1$, such as the terminal double bonded S in a QT (S=P($S_{1/2}$)$_3$) unit, one need count only ½ a bond stretching constraint as discussed elsewhere [41,42,43].

Our Raman scattering results (Fig.4a-d) reveal that the local structures present in the intermediate phase of the present glasses include CS-GeS$_4$, ES-GeS$_2$, PYR-P($S_{1/2}$)$_3$, QT-S=P($S_{1/2}$)$_3$. A count of *bond-stretching* and *bond-bending* constraints per atom for each of these local structures shows (Table 3) a value of 3, exactly equal to the degrees of freedom of an atom in a 3d network. Thus, constraint counting algorithms demonstrate that these local structures are *isostatically rigid*. Furthermore, numerical simulations [44] also suggest that for these local structures to be part of a globally isostatically rigid backbone, they must form n-membered rings with n ~ 6. At present we are unable to address this aspect of structure from the local probe results at hand. However, it is possible that diffraction measurements [45] coupled with numerical simulations [46-48] may permit addressing this key issue of structure in the future. The width of the intermediate phase is broadly related to the range of chemical stoichiometries ($2.28 < \bar{r} < 2.67$) spanned by these local structures (Fig. 5). In the present ternary, the mean coordination number $\bar{r} = 2 + 3x$, and the intermediate phase of $0.090 < x < 0.135$ translates into the range $2.270 < \bar{r} < 2.405$, yielding a width of $\Delta \bar{r} = 0.135(15)$ in reasonable accord with these local structures.

An important feature of the intermediate phase glass compositions is that the non-reversing enthalpy is not only minuscule but that term apparently does not evolve in time, i.e., the glass network *does not age*. The backbone by virtue of an optimal connectedness exists in a state of equilibrium, and it appears that a consequence of that equilibrium is that these glasses do not age. The presence of a small but finite concentration of $P_4S_{10}$ molecules (Fig. 6) in the intermediate phase attests to the robustness of the latter.

Recent Raman scattering experiments [49] on binary Ge-Se glasses performed as a function of hydrostatic pressure provide independent evidence that intermediate phase glass compositions exist in a state of equilibrium. The observation is related to the vanishing of the pressure threshold, $P_c$, in the intermediate phase but its presence in both floppy and stressed-rigid glass compositions. In these experiments, $P_c$ serves as a measure of network internal stress that must be exceeded by an external pressure for the vibrational modes of a network to compress and blue-shift. The nature of medium range structures prevailing in such glasses that produce this remarkable equilibrium in a disordered network is the subject of ongoing investigations.

The intermediate phase in the Ge-P-Se ternary spans a range of $2.255 < \bar{r} < 2.465$, yielding a width [34] of $\Delta \bar{r} = 0.210(15)$, which is significantly larger than the one in the present sulfide glasses. The result is all the more interesting because the nature of the isostatic P- and Ge- centered local structures that contribute to self-organization in both ternaries share much commonality. There are, however, some subtle differences of structure as well. For example, the $P_4S_{10}$ and $P_4S_7$ monomers known to exist in the sulfides, are apparently unstable with Se anions, we suppose, because of the lower overlap of P d-orbitals with Se p-orbitals lowering the strength of the double bond. $P_4S_{10}$ monomers persist in the intermediate phase (Fig.6) at a low level, and are rapidly converted to their S-deficient counterparts, $P_4S_7$ monomers, in the $0.15 < x < 0.18$ range as glasses become increasingly S-deficient. Furthermore, the appearance of P-P containing monomers, $P_4S_4$ and $P_4S_3$ just above the window (Fig.6), suggests that the network backbone must become depleted of the 4 isostatically rigid local structures as these monomers evolve once $x > 0.13$. We believe these structure-related factors contribute to a down-shift of $x_c(2)$, the upper end of the window, from a value of 0.155(5) in the selenide to 0.135 (5) in the corresponding sulfide glasses.

On the other hand, the lower edge of the intermediate phase is up-shifted from $x_c(1)$ = 0.085(5) in the $Ge_xP_xSe_{1-2x}$ ternary to $x_c(2)$ = 0.09(5) in the present $Ge_xP_xS_{1-2x}$ ternary. Although the shift in the threshold is small, it is systematic and unmistakable (Fig.7b). Its molecular origin is tied to the presence of $S_8$ rings that prevail at $x < 0.08$ and preclude a backbone to form let alone become rigid. In the selenide glasses almost all the Se occurs in a polymeric form that promotes forming the backbone even at $x < 0.08$.

*3.2.3 Stressed-Rigid Phase (x > 0.135)*

In the present glasses, the mode at 340 cm$^{-1}$ has contributions from vibrational modes of several species including the CS tetrahedra of Ge(S$_{1/2}$)$_4$ units. Some of these other species include P$_4$S$_7$ (346 cm$^{-1}$), P$_4$S$_3$ (342 cm$^{-1}$), and P$_4$S$_4$ (340 cm$^{-1}$), and each of these units contribute to the scattering near the 345 cm$^{-1}$ mode once x > 0.14. For these reasons, it is not possible to infer the optical elasticity power-law [2,3,34] in the *stressed-rigid* glasses from a measurement of the 345 cm$^{-1}$ mode frequency as a function of glass composition.

Important insights into the thermal behavior of glasses in the *stressed-rigid* phase have emerged from the MDSC measurements [50]. The shape of the non-reversing heat flow near T$_g$ in stressed-rigid glass compositions differs significantly from the one observed in floppy glass compositions; the exotherm is not only broad (> 50 °C) but also has an asymmetrical shape- a tail that extends to temperatures well above T$_g$. It appears that structural arrest in such systems occurs at temperature well above T$_g$, we suppose because the network is mechanically over-constrained. In this composition range, the backbone consists of corner- sharing Ge(S$_{1/2}$)$_4$ tetrahedral units that are overconstrained in that they possess a count of bonding constraints of 3.67 per atom (Table 3) that exceeds the threshold value of 3 coming from the 3d nature of the backbone.

Our results in the *stressed-rigid* glass compositions are limited because we were unable to obtain homogeneous glasses at x > 0.19 in the present sulfides. The result is not entirely unexpected given the proliferation of molecular phase separation discussed above. But in corresponding ternary selenides [34] where such effects are qualitatively absent, many more glass compositions could be synthesized and examined by Raman scattering.

These results show that modes of tetrahedral $P_4$ and amorphous P evolve at higher x in the selenides. And one expects a qualitatively similar behavior to occur in the present sulfides.

**4. Nanoscale Phase Separation and Self-organization effects in ternary Ge-P-S glasses.**

The Ge-P-S ternary displays an extensive glass forming region sketched in Fig. 8 taken from [51]. The glass compositions studied in the present work reside along the bisector of the P-S-Ge angle in the composition triangle. Thermal behavior and molecular structure of binary $P_xS_{1-x}$ glasses were examined in NMR, Raman scattering and DSC measurements by H.Eckert et al. [22,23] who noted the existence of a global maximum in $T_g$ near x ~ 0.20 (Fig.9). Their NMR results showed binary glass compositions at x > 0.20 to be composed of $P_4S_n$ monomers with n = 7,10, i.e. nanoscale phase separated. Glass compositions along the $P_2S_5$- $GeS_2$ join (shown by brown open squares in Fig. 8) were examined in $^{31}$P NMR, Raman scattering and DSC measurements by B. R. Cherry et al. [52] In a separate study, the same authors examined ternary $Ge_{0.714x}P_{0.286x}S_{1-x}$ glasses (open blue triangles in Fig. 8) in $^{31}$P NMR, Raman scattering, DSC and neutron elastic measurements [53]. These latter glasses differ from the one examined presently in that the Ge/P cation ratio is 2.5 instead of 1. Of interest is the fact that glass transition temperatures, $T_g(x)$, increase with x to show a threshold near x = 1/3 (Fig.9) and to decrease thereafter ( x > 1/3). The behavior is reminiscent of the one encountered in binary $Ge_xS_{1-x}$ glasses [54] near the chemical threshold of $x_t$ = 1/3, which constitutes evidence for nanoscale phase separation [38] of a Ge-rich phase ( containing Ge-Ge bonds) from the $Ge(S_{1/2})_4$ backbone once x > $x_t$. These nanoscale phase separation effects can be qualitatively suppressed when the Ge/P ratio equals 1 as reflected in the monotonic increase of $T_g(x)$ observed in the present glasses (Fig 9). The pattern is not unique to the present Ge-P-S ternary but has been observed in other ternaries such as Ge-P-Se, Ge-As-Se, Ge-As-S as discussed elsewhere [38]. Spectroscopic probes such as Raman scattering, NMR, Mossbauer effect, provide powerful probes of local structure in glasses. However, from a measurement at one glass composition it is generally impossible to tell if these local structures form part of the same backbone or are formed in separate molecular clusters or backbones. In this latter respect, variations in $T_g$ with glass composition have turned out to be decisive as discussed elsewhere [38,54]. The intrinsically nanoscale phase separated regions are shown with hashed lines in Fig. 8.

The intermediate phase established in the present ternary glasses along with our tentative estimate of this phase in the Ge-S binary is sketched green in Fig. 8. The intermediate phase in Ge-S glasses has not been firmly established, although its light-induced collapse at the mean-field rigidity transition has been observed [4]. Currently, there is no information on the existence of the intermediate phase in the P-S binary.

One cannot help but comment on the robustness of the intermediate phase in the present $Ge_xP_xS_{1-2x}$ ternary glasses. In the intermediate phase, we observe evidence of the 4 isostatic local structures that form the building blocks of the stress-free backbone eluded to earlier. In addition, we also observe evidence of a small (< 5 mole %) but finite concentration of $P_4S_{10}$ monomers in this phase. And apparently, this small concentration of monomers does not influence the non-aging behavior of the intermediate phase up to 8 weeks. It is of interest to examine the location of the intermediate phase in the P-S and Ge-S binaries unambiguously and to establish the phase boundaries of the three elastic phases in these ternary glasses as was done in corresponding selenide glasses [2].

## 5. Conclusions

In the present work, we have examined ternary $Ge_xP_xS_{1-2x}$ glasses over a wide composition range, $0.05 < x < 0.19$, using Raman scattering and MDSC experiments. The latter experiments show the existence of a *thermally reversing window* in the $0.09 < x < 0.135$ range. The observation suggests that glasses at $x < 0.09$ are elastically floppy, those in the $0.09 < x < 0.135$ in the intermediate phase while those at $x > 0.135$ are stressed-rigid. Furthermore, the MDSC experiments reveal that for glass compositions in the intermediate phase, not only is the non-reversing heat flow term, $\Delta H_{nr}$, minuscule (~0) but that term does not age for a waiting time of 8weeks. In stressed-rigid and floppy glass compositions, the $\Delta H_{nr}$ term is not only an order of magnitude larger but the term is found to age. Raman scattering experiments show that there are four isostatic local structures including CS-$GeS_4$, ES-$GeS_2$, PYR-$P(S_{1/2})_3$, QT-S=$P(S_{1/2})_3$ populated in the intermediate phase. The width of the intermediate phase in the present ternary sulfides of $\Delta \bar{r} = 0.135(15)$ is significantly narrower than in corresponding selenides $\Delta \bar{r} = 0.210(15)$. The narrower intermediate phase width in the sulfides in relation to the selenide glasses is ascribed to the presence of nanoscale phase separation of $S_8$ monomers in the floppy phase, and $P_4S_4$ and $P_4S_7$ monomers in the stressed-rigid phase that suppresses formation of a backbone in which rigidity could percolate.

**Acknowledgments**

We have benefited from discussions with D. McDaniel, B.Goodman, and L. Koudelka, during the course of this work. This work was supported by NSF grant DMR-01-01808.20

[14]   Georgiev D G, Boolchand P, Eckert H, Micoulaut M and Jackson A 2003 *Europhysics Letters* **62** 49

[15]   Ward A T 1968 *J. Phys. Chem.* **72** 4133

[16]   Qu T and Boolchand P cond-mat/0312481 at http://www.arxiv.org

[17]   *Modulated DSC$^{TM}$ Compendium* 1997 Reprint #TA-210 (New Castle, DE: TA Instruments Inc) http://www.tainst.com/

[18]   Lucovsky G, Galeener F L, Keezer R C, Geils R H and Six H A 1974 *Phys. Rev.* B **10** 5134

[19]   Boolchand P, Grothaus J, Tenhover M, Hazle M A and Grassell R K 1986 *Phys. Rev.* B **33** 5421

[20]   Koudleka L and Pisarcik M 1990 *Phys. Chem. Glasses* **31** 217

[21]   Koudleka L, Pisarcik M, Gutenev M S and Blinov L N 1989 *J. Mat. Sci. Lett.* **8** 933

[22]   Shibao R K, Xia Y, Srdanov V I and Eckert H 1995 *Chem. Mater.* **7** 1631

[23]   Mutolo P, Witschas M, Regelsky G, Schmedt auf der Guenne J and Eckert H 1999 *J. Non-Cryst. Solids* **256-257** 63

[24]   Chattopadhyay T, Carlone C, Jayaraman A and Schnering H G V 1981 *Phys. Rev.* B **23** 2471

[25]   Gardner M 1973 *J. Chem. Soc. Dalton* **6** 691

[26]   Jensen J O and Zeroka D 1999 *J Mol. Struct. (Theochem)* **487** 267

[27]   Lannin J S and Shanabrook B V 1978 *Solid State Commun.* **28** 497

[28]   Andreas K, Alexander K and Martin T 2002 *J. Chem. Phys.* **116** 3323

[29]   Jackson K, Briley A, Grossman S, Porezag D V and Pederson M R 1999 *Phys. Rev.* B **60** R14985
22

# Table 1

| Mode Frequency (cm$^{-1}$) | Mode Labels | Local Structures | References |
|---|---|---|---|
| 150, 218, 475 | S$_n$ | S-chains and Rings | Lucovsky G, Galeener F L, Keezer R C, Geils R H and Six H A 1974 *Phys. Rev.* B **10** 5134 |
| 185, 286, 342, 421, 441, 488 | P$_4$S$_3$$^m$ | P$_4$S$_3$ | Chattopadhyay T, Carlone C, Jayaraman A and Schnering H G V 1981 *Phys. Rev.* B **23** 2471 |
| 189, 217, 232, 303, 346, 444 | P$_4$S$_7$$^m$ | P$_4$S$_7$ | Gardner M 1973 *J. Chem. Soc. Dalton* **6** 691 |
| 196, 270, 715 | P$_4$S$_{10}$$^m$ | P$_4$S$_{10}$ | Jensen J O and Zeroka D 1999 *J Mol. Struct. (Theochem)* **487** 267 |
| 243, 285, 440, 462 | P$_4$S$_4$$^m$ | P$_4$S$_4$ | Ystenes M, Menzel F and Brockner W 1994 *Spectrochimica Acta.* **50a** 225 |
| 346 | CS | Corner sharing GeS$_4$ | Jackson K, Briley A, Grossman S, Porezag D V and Pederson M R 1999 *Phys. Rev.* B **60** R14985 and Feng X, Bresser W.J and Boolchand P 1997 *Phys. Rev. Lett.* **78** 4422 |
| 375 | ES | Edge Sharing GeS$_2$ | |
| 389, 398, 425, 458, 500 | a-P | Amorphous P | Lannin J S and Shanabrook B V 1978 *Solid State Commun.* **28** 497 |
| 416 | PYR | P(S$_{1/2}$)$_3$ | Koudleka L and Pisarcik M 1990 *Phys. Chem. Glasses* **31** 217 |
| 440 | F$_2$ | Face Sharing GeS$_{4/2}$ | Feng X, Bresser W.J and Boolchand P 1997 *Phys. Rev. Lett.* **78** 4422 |
| 465, 606 | P$_4$ | P$_4$ clusters | Andreas K, Alexander K and Martin T 2002 *J. Chem. Phys.* **116** 3323 |
| 700 | QT | S=P(S$_{1/2}$)$_3$ | Koudleka L and Pisarcik M 1990 *Phys. Chem. Glasses* **31** 217 and Shibao R K, Xia Y, Srdanov V I and Eckert H 1995 *Chem. Mater.* **7** 1631 |

## Table 2

| Bond type | Pauling Single Bond Strength |
|---|---|
| P-P | 51.3 kcal/mole |
| Ge-Ge | 37.6 kcal/mole |
| S-S | 50.9 kcal/mole |
| P-S | 54.8 kcal/mole |
| P-Se | 51.72 kcal/mole |
| Ge-S | 55.52 kcal/mole |
| Ge-Se | 49.08 kcal/mole |

## Table 3

| Unit | Mean Co-ordination Number ($\bar{r}$) | No. of constraints /atom ($n_c$) | Comments |
|---|---|---|---|
| $S_8$ | 2.00 | 2.00 | Monomers; decouple from backbone. Presence inferred from their opening at the $T_\lambda$ transition in MDSC scans. |
| QT $S=P(S_{1/2})_3$ | 2.29 | 3.00 | Isostatic molecular units whose concentration peaks at $x = x_c(2) = 13\%$ |
| $P_4S_{10}$ | 2.29 | 3.00 | Monomeric units present in the glasses beginning at very low values of x and highly populated around $x = 15\%$ |
| $P_4S_7$ | 2.36 | 3.09 | Decoupled monomers appearing in glasses where x>14% |
| PYR $P(S_{1/2})_3$ | 2.40 | 3.00 | Optimally constrained units forming part of backbone. |
| CS $GeS_4$ | 2.40 | 3.00 | Isostatic local structure forming part of the backbone. Their concentration is the highest when $\bar{r}$ = 2.40 or x about 14% |
| $P_4S_4$ | 2.50 | 3.25 | Decoupled monomers; They begin to appear in glasses when x>15% |
| $P_4S_3$ | 2.57 | 3.43 | |
| ES $GeS_2$ | 2.67 | 3.00 | Isostatic local structures forming part of the backbone. |
| CS $Ge(S_{1/2})_4$ | 2.67 | 3.67 | Stressed-rigid units present at x > 13% in glasses |

**Captions**

**Figure 1.** MDSC scans of $Ge_xP_xS_{100-2x}$ glasses at (a) x = 5; floppy glass (b) x = 11; intermediate phase glass and (c) x = 14 ; stressed rigid glass, each taken at 3 ºC/min scan rate, 1ºC/100s modulation rate. The total (green), reversing (blue) and non-reversing (red) heat flow terms are shown. The later scanning up in T ( right arrow) and then down ( left arrow) are illustrated in (b) and (c). The frequency corrected $\Delta H_{nr} = \Delta H_{nr}^{up} - \Delta H_{nr}^{down}$ nearly vanishes for a glass at x = 11 shown in (b). See text for details.

**Figure 2.** Normalized non-reversing term associated with $T_\lambda$ endotherm, $\Delta H_{nr}(T_\lambda)$, showing a steadily decreasing value with x in the 5% < x < 8%. The data reveal a two-fold reduction in $S_8$ concentration across the indicated range of x.

**Figure 3.** MDSC results on present glasses showing trends in (a) $T_g(x)$ and (b) $\Delta H_{nr}(x)$ as a function of x and aging. Filled triangles give the $T_\lambda$ transition observed in S-rich glasses. A sharply defined square-well like thermally reversing window is observed in the 8 week aged samples in (b) while $T_g$s vary monotonically across this window in (a).

**Figure 4.** (a) Observed Raman scattering lineshapes as a function of x with principal modes labeled. (b) and (c) give a blow up of insets shown in (a). Sharp modes of P-rich monomers $P_4S_3$, $P_4S_4$, $P_4S_7$ and $P_4S_{10}$ and QT units are labeled in (b) and (c). (d) shows an example of a deconvolution of a Raman lineshape at x = 15% .

**Figure 5.** Ball and stick diagrams of (a) monomeric molecular units and (b) Isostatic network forming units observed in the present terary glasses. Mean co-ordination number *r* and the count of constraints per atom ($n_c$) are also shown. Also see Table 2.

**Figure 6.** Raman scattering strengths of various vibrational modes of specific molecular units in ternary Ge-P-S glasses plotted as a function of x. The CS mode strength could not be reliably ascertained at x > 13% because of overlap with modes of other units. Note that the scattering strength of modes of the monomers, $P_4S_{10}$, $P_4S_4$, $P_4S_7$ and $P_4S_3$ are scaled up by a factor of 5 to display on the same y-axis scale.

**Figure 7.** Compositional trends in (a) Glass transition temperature ( $T_g$) and (b) Non-reversing heat flow ($\Delta H_{nr}$) in $Ge_xP_xS_{1-2x}$ glasses and corresponding $Ge_xP_xSe_{1-2x}$ glasses compared. The intermediate phase in the sulfide glasses narrows in relation to the selenides. The work on selenide glasses is taken from ref.34.

**Figure 8.** Glass forming region in the Ge-P-S ternary depicting the glass composition studied presently (open green circles). Blue triangle and brown square compositions were studied in ref 52 and 53 respectively. The green shaded region indicates glasses that are self-organized while the red-hashed region glass compositions that are nanoscale phase separated.

**Figure 9.** Compositional trends in glass transition temperatures in $Ge_xS_{1-x}$, $P_xS_{1-x}$, $Ge_xP_xS_{1-x}$ and $Ge_{0.714x}P_{0.286x}S_{1-x}$ glasses plotted as a function of mean co-ordination

number *r*. The arrows mark the mean co-ordination number r at which the chemical threshold $x_t$ occurs in these glass systems.

**Table 1.** Summary of vibrational mode assignments of various molecular units present in the titled glasses.

**Table 2.** Chemical bond strengths of several heteropolar and homopolar bonds among Ge, P, S and Se taken from ref.32.

**Table 3.** Mean co-ordination number $\bar{r}$, and number of constraints per atom $n_c$, of the network forming- and monomeric- units present in Ge-P-S glasses.

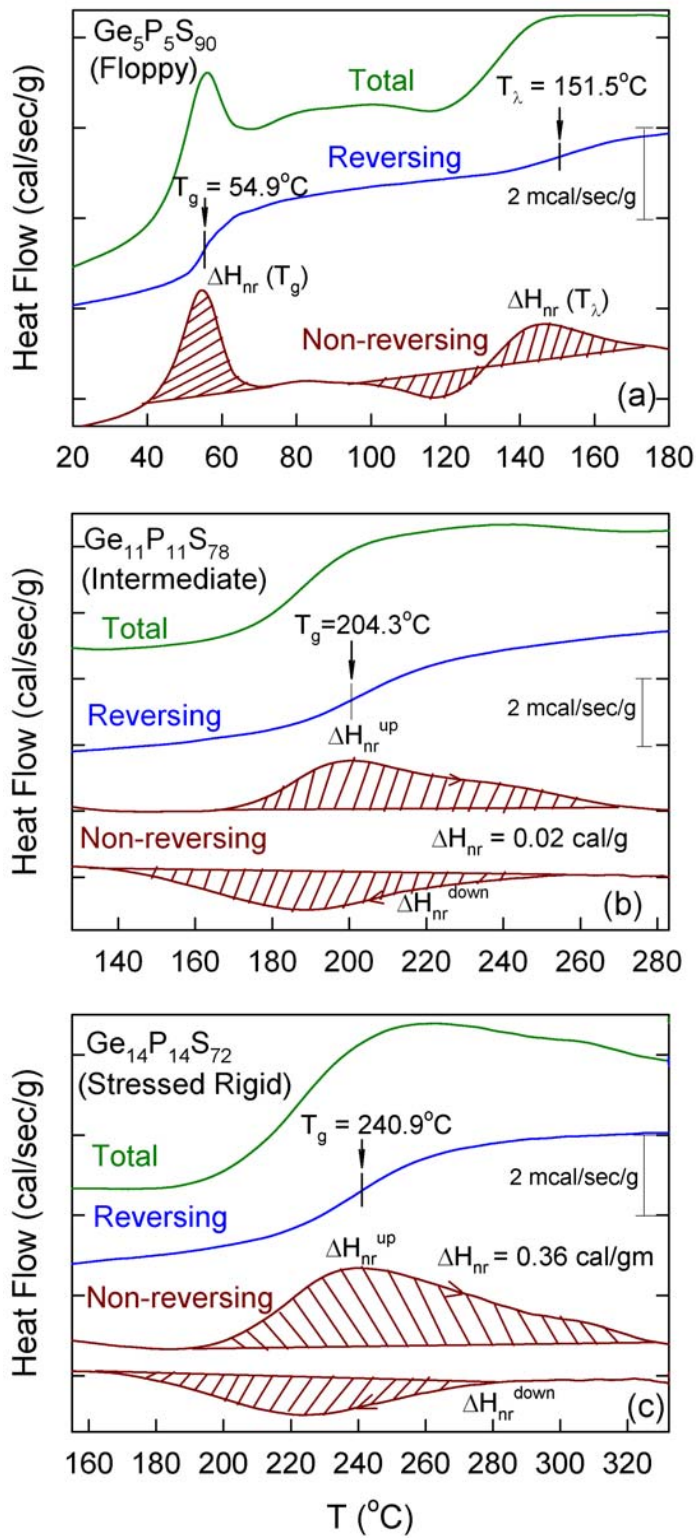

Figure 1

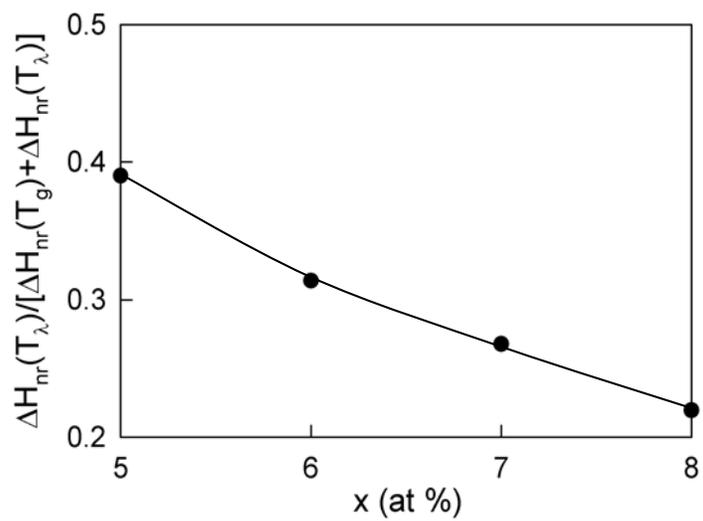

Figure 2

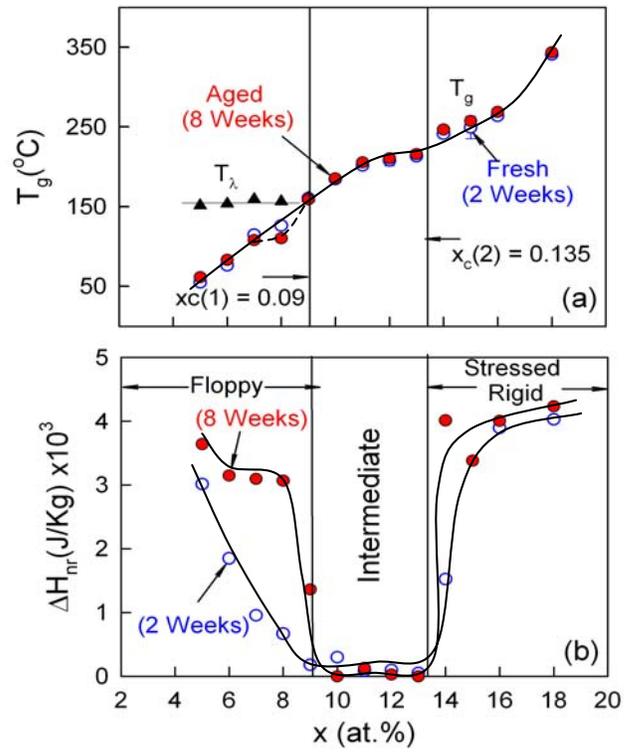

Figure 3

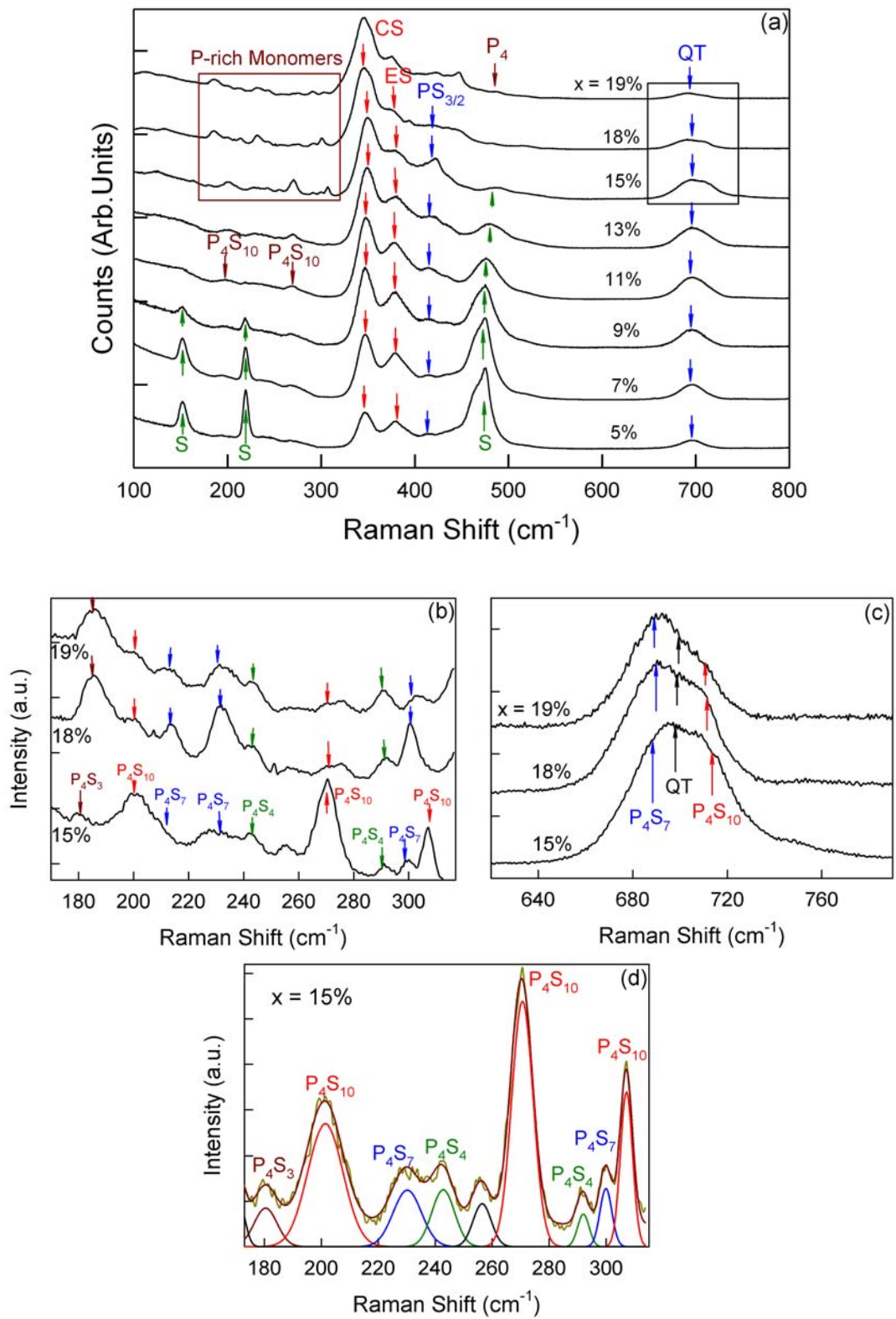

Figure 4

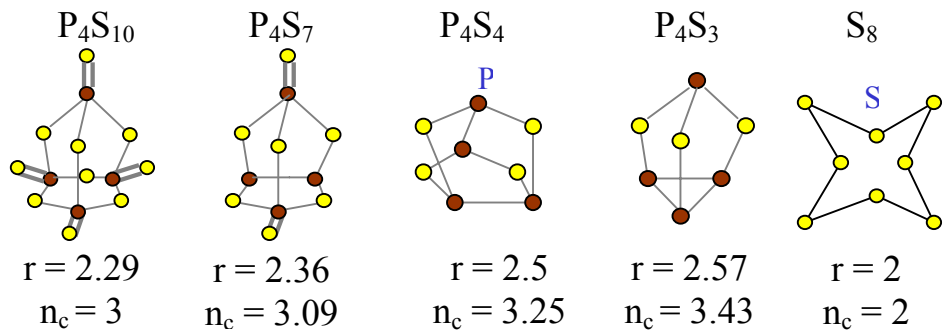

(a) Monomeric Units

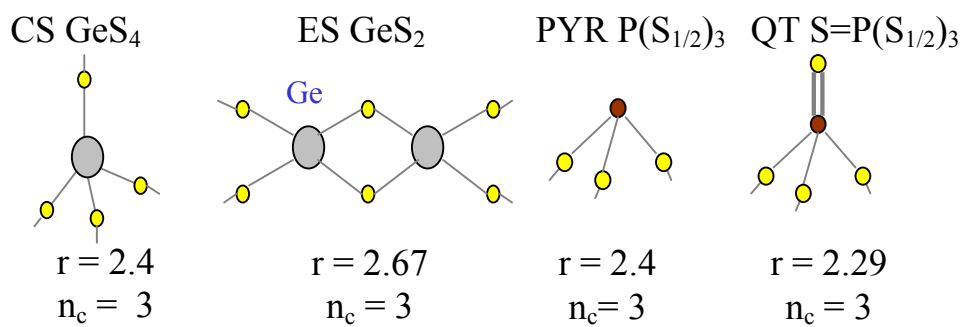

(b) Network Forming Units

Figure 5

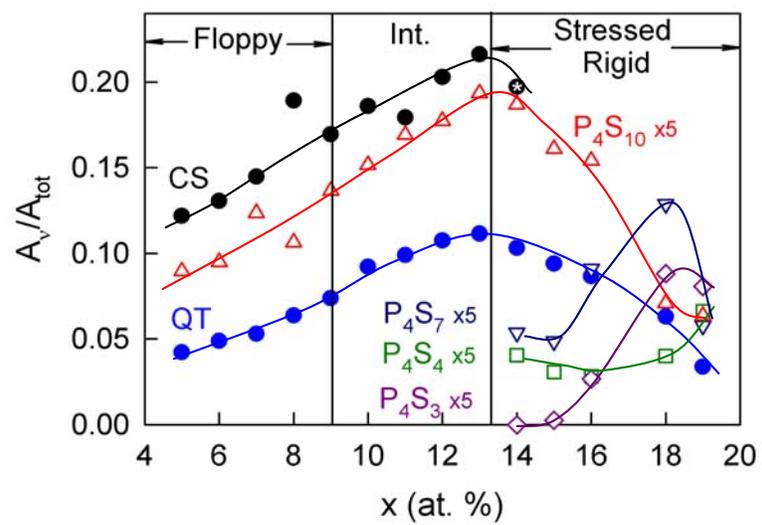

Figure 6

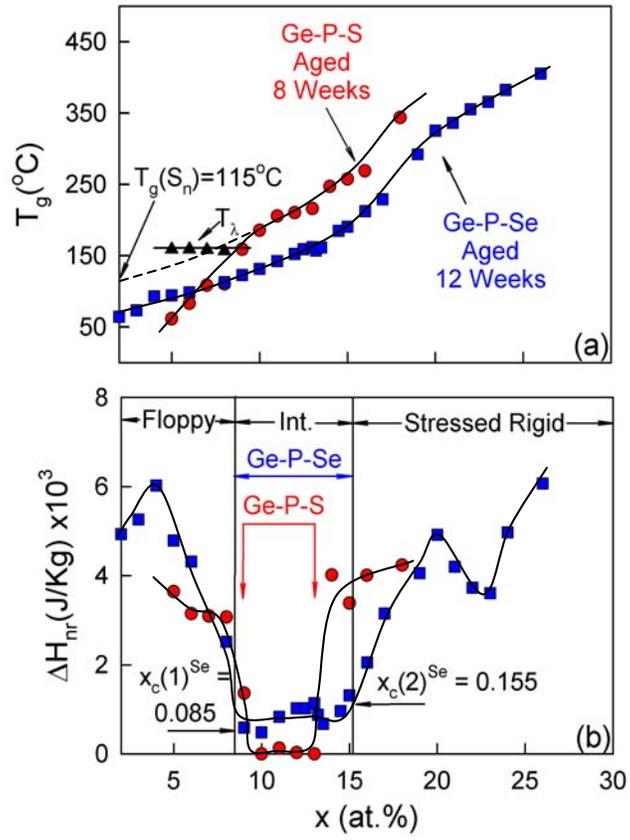

Figure 7

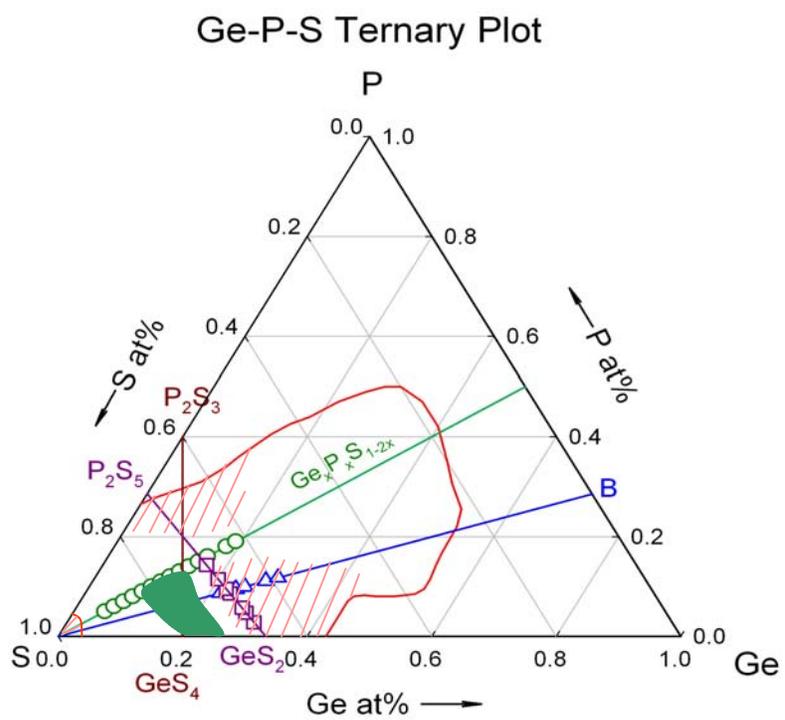

Figure 8

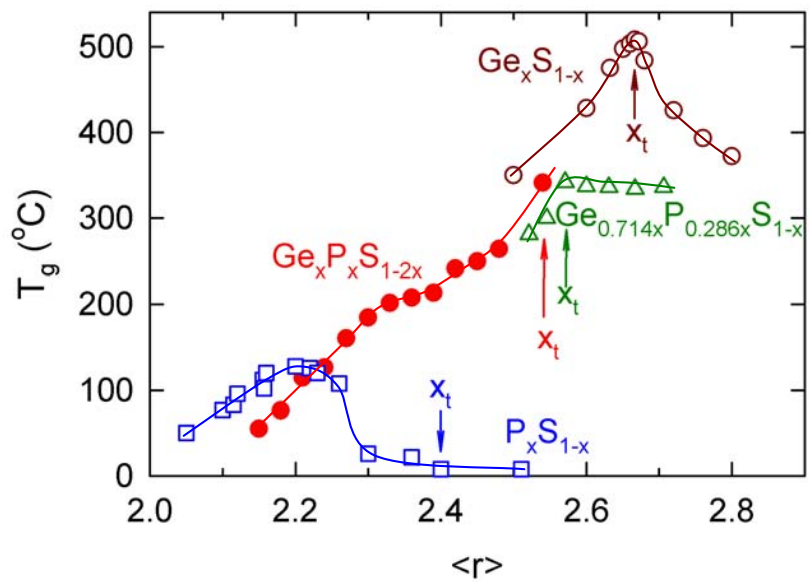

Figure 9